\documentclass[11pt]{article}
\usepackage[english]{babel}
\usepackage{amsmath}
\def\lsim{\,\lower2truept\hbox{${< \atop\hbox{\raise4truept\hbox{$\sim$}}}$}\,}
\def\gsim{\,\lower2truept\hbox{${> \atop\hbox{\raise4truept\hbox{$\sim$}}}$}\,}

\def\deg{\ifmmode^\circ \else$^\circ $\fi}    
\def\arcs{\ifmmode {'' }\else $'' $\fi}     
\def\arcm{\ifmmode {' }\else $' $\fi}     
\def\buildrel#1\over#2{\mathrel{\mathop{\null#2}\limits^{#1}}}
\def\mper{\ifmmode \buildrel m\over . \else $\buildrel m\over .$\fi}
\def\hper{\ifmmode \rlap.^{h}\else $\rlap{.}^h$\fi}
\def\sper{\ifmmode \rlap.^{s}\else $\rlap{.}^s$\fi}
\def\arcsper{\ifmmode \rlap.{' }\else $\rlap{.}' $\fi}
\def\arcmper{\ifmmode \rlap.{'' }\else $\rlap{.}'' $\fi}

\def\mincir{\ \raise -2.truept\hbox{\rlap{\hbox{$\sim$}}\raise5.truept	
    \hbox{$<$}\ }}								%
\def\magcir{\ \raise -2.truept\hbox{\rlap{\hbox{$\sim$}}\raise5.truept	%
    \hbox{$>$}\ }}								%

\def\deg{\ifmmode^\circ \else$^\circ $\fi}    
\def\arcs{\ifmmode {'' }\else $'' $\fi}     
\def\arcm{\ifmmode {' }\else $' $\fi}     
\def\buildrel#1\over#2{\mathrel{\mathop{\null#2}\limits^{#1}}}
\def\mper{\ifmmode \buildrel m\over . \else $\buildrel m\over .$\fi}
\def\hper{\ifmmode \rlap.^{h}\else $\rlap{.}^h$\fi}
\def\sper{\ifmmode \rlap.^{s}\else $\rlap{.}^s$\fi}
\def\arcsper{\ifmmode \rlap.{' }\else $\rlap{.}' $\fi}
\def\arcmper{\ifmmode \rlap.{'' }\else $\rlap{.}'' $\fi}

\def\mincir{\ \raise -2.truept\hbox{\rlap{\hbox{$\sim$}}\raise5.truept	
    \hbox{$<$}\ }}								%
\def\magcir{\ \raise -2.truept\hbox{\rlap{\hbox{$\sim$}}\raise5.truept	%
    \hbox{$>$}\ }}								%

\begin {document}

\null

\bigskip
\bigskip
\bigskip
\bigskip
\bigskip
\bigskip
\bigskip
\bigskip

\hspace{2cm}
{\small  Internal Report ITESRE 198/1997}

\bigskip
\bigskip

\hspace{2cm} {\bf A PRELIMINARY STUDY ON DESTRIPING}

\hspace{2cm} {\bf TECHNIQUES OF PLANCK LFI MEASUREMENTS}

\hspace{2cm} {\bf VERSUS OBSERVATIONAL STRATEGY}

\medskip

\hspace{2cm}
{\sc C.~Burigana$^1$, M.~Malaspina$^1$, N.~Mandolesi$^1$,}

\hspace{2cm}
{\sc L.~Danese$^2$, D.~Maino$^2$,}

\hspace{2cm}
{\sc M.~Bersanelli$^{3,1}$ \& M.~Maltoni$^{4,1}$}

\bigskip
\bigskip

\hspace{2cm}
$^1${\it Istituto TESRE, CNR, Bologna, Italy}

\hspace{2cm}
$^2${\it SISSA -- International School for Advanced Studies,
  Trieste, Italy}

\hspace{2cm}
$^3${\it Istituto di Fisica Cosmica, CNR, Milano, Italy}

\hspace{2cm}
$^4${\it Dipartimento di Fisica, Universit\`a di Ferrara, Italy}

\bigskip
\bigskip
\bigskip
\bigskip

\hspace{2cm}
November 1997

\bigskip
\bigskip
\bigskip
\bigskip

\newpage

\begin{center}

{\small  Internal Report ITESRE 198/1997}

\bigskip
\bigskip
\bigskip
\bigskip
\bigskip
\bigskip

\large

{\bf A PRELIMINARY STUDY ON DESTRIPING}

{\bf TECHNIQUES OF PLANCK LFI MEASUREMENTS}

{\bf VERSUS OBSERVATIONAL STRATEGY}

\bigskip
\bigskip

\normalsize

\bigskip

C.~Burigana$^1$, M.~Malaspina$^1$, N.~Mandolesi$^1$,
L.~Danese$^2$, D.~Maino$^2$, M.~Bersanelli$^3$ \& M.~Maltoni$^{4,1}$

\bigskip

$^1${\it Istituto TESRE, CNR, Bologna, Italy}

$^2${\it SISSA -- International School for Advanced Studies,
  Trieste, Italy}

$^3${\it Istituto di Fisica Cosmica, CNR, Milano, Italy}

$^4${\it Dipartimento di Fisica, Universit\`a di Ferrara, Italy}

\bigskip
\bigskip
\bigskip
\bigskip
\bigskip
\bigskip

\end{center}

\begin{quotation}

SUMMARY -- We present here the basic issues of our simulations
of Planck observations, 
focusing on the study of the stripes generated
by $1/f$-type noise due to amplifiers noise temperature variation
of LFI radiometers generated by gain fluctuations.
Our simulations include a realistic estimate of $1/f$ knee frequency,
based on
a recent analytical study of the properties of the Planck LFI
radiometers and of their systematic effects,
as well as realistic choices of load, amplifier noise
and paylod environment temperatures.
By comparing simulated  observed maps of CMB anisotropies
obtained by including or not the instrumental noises
we are able to quantify the magnitude of the striping effect.
The ``standard'' destriping method presented
in the report on the Phase A study has been adapted to our purposes.
We quantify the efficiency
of this destriping technique for different scanning strategies,
for typical off-axis LFI beams. Our results are compared with those of
other previous analyses.

\end{quotation}

\section{Introduction} \label{sec_10}

The Planck Surveyor is an European Space Agency (ESA) satellite mission to map
spatial anisotropy in the Cosmic Microwave Background (CMB) over a wide
range of frequencies with an unprecedented combination of sensitivity,
angular resolution, and sky coverage (Bersanelli et al. 1996).
The data gathered by this mission will revolutionize modern cosmology by
shedding light on fundamental cosmological questions such as the age and
present expansion rate of the universe, the average density of the universe,
the amount and the kind of dark matter, and other questions.
As with any CMB experiment, achieving the desired performance requires careful
attention to the control of systematic effects.
$1/f$-type noise in the radiometer output is
one of the most critical systematic effect pertaining to the
Low Frequency Instrument (LFI) radiometers
because it may lead
to striping in the final sky maps and increase the noise level. In general,
a value of the $1/f$ knee frequency, $f_k$, significantly
greater than the spacecraft
rotation frequency, $f_s$, will lead to some degradation in sensitivity
(Janssen et al. 1996). 
In this paper we examine
the impact of the $1/f$-type noise
by adopting a realistic estimate of $1/f$ knee frequency
as a function of the load, amplifier noise
and paylod environment temperatures, of the radiometer bandwidth
and of the level of fluctuations of each stage of the radiometer amplifiers,
based on
recent analytical studies of systematic effects in Planck LFI
radiometers (Seiffert et al. 1997).

In section~\ref{sec_20} we summarize the basic concepts relevant for 
understanding
the basic properties of instrumental noises.

In section~\ref{sec_30} we present the
mathematical formalism of our numerical code for the
simulation of the Planck observations, including instrumental noises,
and the data stream generation; we typically
refer here to Planck like scanning
strategies, but our code is versatile enough
to allow the study of other observational schemes.

In section~\ref{sec_40} we describe how we have converted 
our simulated data streams in simulated observed maps.

In section~\ref{sec_50} we discuss in detail the mathematical formalism of
the proposed destriping technique, including some considerations
about the numerical efficiency.

The ``standard'' estimators that quantify the magnitude of the
striping effect and the efficiency of destriping techniques are presented
in section~\ref{sec_60}.

Some preliminary results are presented in section~\ref{sec_70}.

Finally, in section 8 we draw out the main conclusions of our
analysis, compare our results with those of previous works
and draw out a brief guide-line for a future work. We discuss there
the main implications of our study,
focusing on their impact for the optimization
of the Planck observational strategy.

\section{Sources of instrumental noises} \label{sec_20}

Planck LFI radiometers are
modified Blum correlation receivers (Blum 1959, Colvin 1961).
The modification is that the temperature of the reference load is quite
different from the sky temperature (Bersanelli et al. 1995). To compensate,
differing DC gains are applied after the two detector diodes.
Adjusting the ratio of DC gains, $r$, allows one to null the output signal,
minimize sensitivity to RF gain fluctuations, and
achieve the lowest white noise in the output.
Although it may not be immediately apparent, the fact that the reference
load is not at the same temperature as the sky does not increase the
white noise level compared to a standard correlation receiver.

The ideal sensitivity of our radiometer for a single observation
with an integration time $\tau$ is

\begin{equation} \label{eq_10}
  \Delta T_{\rm white} =
  \frac{\sqrt{2} \left({T_n + T_x}\right)}{\sqrt{\beta \tau}} \, ,
\end{equation}

\noindent
where $\beta$ is the effective bandwidth,
$T_x$ is the noise temperature of the signal entering one of the two horns
and $T_n$ is the amplifier noise temperature.

In order to null the average output signal of the radiometer,
the DC gains ratio after the two detector diodes, $r$,
must be adjusted to the proper value.
In the simple case that the gains of the two amplifiers entering the
two horns and their noise temperatures can be considered equal,
$r$ must be setted to the value:

\begin{equation}
  r = \frac{T_x + T_n}{T_y + T_n} \, ,
\end{equation}
where $T_y$ is the reference load temperature entering the other horn.
The above temperatures are antenna temperatures;
$T_x$ is due properly to the sum of the sky temperature (essentially the
CMB monopole antenna
temperature, related to the CMB thermodynamic temperature $T_0 \simeq 2.726$K,
plus ``minor'' contributions from
CMB dipole and anisotropies, galactic and extragalactic foregrounds,
bright sources, Zodiacal light ...) and of the ``environment'' temperature
(of about 1 K) due to the satellite emission.

There are several potential concerns for the current radiometer scheme.
Amplifiers noise temperature variations, that derive from gain fluctuations,
could be confused with the sky signal variations, that
we are interested in measuring, by introducing
a change in the observed signal, $\Delta T_{\rm equiv}$, which can mimic a
true sky fluctuation.
The amplifiers noise temperature variations
have the characteristics of $1/f$ noise and
this leads to $1/f$-type noise in the radiometer output.

We have recently estimated the $1/f$ knee frequency 
of our radiometer (Burigana et al. 1997a, Mandolesi et al. 1997a,
Seiffert et al. 1997)
on the simplifying assumption that there is no $1/f$ contribution from
the detector diodes and neglecting the effects of the phase
shifting that is designed to control this contribution;
the diodes are assumed to be
perfect square law detectors
and we have assumed that the bandpass of the signal is the same
in both legs of the radiometer.
Here we briefly summarize 
the main arguments relevant
for the expected magnitude
of gain and noise temperature fluctuations and 
the main concepts relevant for the estimate of the
$1/f$ knee frequency.

The contribution to the $1/f$ noise
directly due to amplifier gain fluctuations results to be
zero at first order.
Under quite reasonable assumptions,
the noise contributions due to reference load fluctuations and
fluctuations in the ratio of DC gains are much less than that due to
the amplifier noise temperature fluctuations term, which then
results to be the dominant source of $1/f$ noise in the radiometer output.
Imperfect isolation does not significantly modify this conclusion.
Further, the sensitivity of our
radiometer to differences between the gains and noise temperatures
of the two amplifiers is not critical.
As a consequence, all these complications cannot 
significantly change the knee frequency respect to 
the results we draw out here below.

Cryogenic HEMT amplifiers have noise temperature fluctuations with
a $1/f$-type spectrum because induced by the $1/f$-type
gain fluctuations of amplifiers (Wollack 1995, Jarosik 1996,
Seiffert et al. 1996).
The magnitude of noise temperature fluctuations can be computed from
the following argument.
Assuming that each stage of the amplifier has the same level of fluctuation,
we can conclude that the transconductance
of an individual HEMT device also fluctuates according to:
\begin{equation}
  \frac{\Delta g_m}{g_m} = \frac{1}{2 \sqrt{N_s}}\frac{\Delta G}{G} \, ,
\end{equation}

\noindent
where $N_s$ is the number of stages of the amplifier, typically $\sim 5$.
An optimal low noise amplifier design will have equal noise contributions from the
gate and drain of the HEMT, which means the changes in $g_m$ will lead to
changes in $T_n$ (Pospieszalsky 1989). This can be expressed as
\begin{equation}
\frac{\Delta T_n}{T_n} \simeq \frac{\Delta g_m}{g_m} \, .
\end{equation}

We can write the $1/f$ spectrum of the gain fluctuations as:
\begin{equation}
  \frac{\Delta G}{G} = \frac{C}{\sqrt{f}} \, .
\end{equation}

Putting this together we get:
\begin{equation}
  \frac{\Delta T_n}{T_n} \simeq \frac{1}{2 \sqrt{N_s}}\frac{C}{\sqrt{f}} \, .
\end{equation}

We can therefore write the amplifier noise temperature fluctuations as
\begin{equation}
  \frac{\Delta T_n}{T_n}  = \frac{A}{\sqrt{f}} \, ,
\end{equation}
with $A=C/(2\sqrt{N_s})$; a normalization of $A \simeq 1.8 \times 10^{-5}$
(relying on the references above) is appropriate for the Planck  radiometers at 30 and 45 GHz.
Throughout, we will use units of $\rm{K} / \sqrt{\rm Hz}$ for $\Delta T$
so that we will not need to refer to the sampling frequency of the radiometer.
In these units then, $\Delta T / T$ has units of Hz$^{-1/2}$ and $A$ is dimensionless.
We also note that the value of $A$ will generally depend on the physical temperature
of the amplifier. The values for $A$ given here should be regarded as estimates
rather than precise values.
For the radiometers at higher
frequencies, it will be
necessary to use HEMT devices with a smaller gate width to achieve the lowest
amplifier noise figure. We expect that the gate widths will be
roughly $1/2$ that of the devices used for the lower frequency radiometers and this will
lead to $g_m$ fluctuations that are roughly a factor of $\sqrt{2}$ higher
(Gaier 1997, Weinreb 1997).  We will therefore adopt a normalization of
$A=2.5 \times 10^{-5}$ for the 70 and 100 GHz radiometers.

Starting from the expression of the average output of the differential
radiometer, one derives the change in the output signal 
for the above small change
in the noise temperature of one of the amplifier;
and by multipling it by a factor
$\sqrt{2}$ because both amplifiers (which have uncorrelated noise)
can contribute to this effect, we have
the change in the observed signal, $\Delta T_{\rm equiv}$, given by:

\begin{equation}
  \Delta T_{\rm equiv} = \sqrt{2} \Delta T_n \left[{ \frac{1 -r}{2} }\right] \, .
\end{equation}

\noindent

We define the ``knee'' frequency as the post-detection frequency,
$f_k$ at which the $1/f$ contribution and the ideal
white noise contribution are equal,
i.e. $\Delta T_{\rm equiv} = \Delta T_{\rm white}$.
For the computation of the knee frequency
we use an integration time $\tau = 1/(2 \Delta f)$  and $\Delta f = 1$ Hz, according to the
choosen units for $A$ and $\Delta T_n$.
The knee frequency is then given by:

\begin{equation} \label{eq_90}
  f_k = \frac{ A^2 \beta}{8} {\left({1-r}\right)}^2 {\left({\frac{T_n}
      {T_n + T_x}}\right)}^2 \, .
\end{equation}

This expression shows as the knee frequency
depends upon several factors including the radiometer bandwidth, reference
load temperature, and the intrinsic level of fluctuation in the HEMT devices;
values of $\sim \mathrm{few} \times 0.1$ Hz, according to the frequency,
should be reached with only passive
cooling of the radiometer (to about 50 K), whereas active cooling
(to about 20K or less) can further reduces the knee frequency.
As examples, assuming a 20\% bandwidth for our frequency channels
and an antenna temperature $T_x = 3$K,
the knee frequency is $0.046$ Hz and
$0.11$ Hz respectively at 30 GHz (assuming $T_y=20$K and $T_n=10$K)
and at 100 GHz (assuming $T_y=20$K and $T_n=40$K).

For a more realistic evaluation we have to accurately repeat 
the analytical considerations by Seiffert et al. (1997)
to distinguish between 
the amplifier noise temperature, $T_n$, and the system temperature,
$T_{sys}$. In reality, it would be more appropriate to insert
$T_{sys}$ in eq. (1) and to carefully consider when the two temperatures
enter in the determination of the knee frequency. On the other hand,
the difference between these two temperature is 
estimated to be of few K while the 
instrumental situation is still partially unclear; then a 
careful distinction between them is presently not necessary
from practical point of view, although interesting.
We will address this point in a future work.

The knee frequency must be compared to the spin frequency $f_s$;
for the Planck observational strategy proposed for the Phase A study
$f_s = 1$ r.p.m., i.e. 0.017 Hz.

As a comparison, for a total power radiometer, laboratory measurements have found
knee frequencies between 10 and 100 Hz; the modified
correlation radiometer scheme reduces the knee frequency by
more than two order magnitudes.

\section{Simulation of the mission} \label{sec_30}

We have written a code that simulates the basic properties of Planck
observations in order to study  the striping effect
due to $1/f$-type noise on the measured sky temperatures.
For simplicity our sky map includes only CMB fluctuations, generated 
in equi-cylindrical pixelisation (ECP) by using the method of 
Muciaccia et al. (1997); then we have projected
it in COBE-cube pixelisation in order to have quasi equal area pixels. 
For our tests we have considered CMB fluctuations in 
a typical CDM scenario ($\Omega_b =0.05$, with COBE/$DMR$ normalization). 
We have 
used here
an input map $\bf M_{in}$ at a resolution of $19.4'$ 
[i.e. at COBE-cube resolution 9]
that we have derived
from an ECP map with 1024 
grid points along each parallel
computed by using the multipoles up $l=512$.
The typical dimension of a pixel
of an input map at resolution 9 is comparable with the beam FWHM of $30'$
of the channels at 30GHz, that we have considered for the present tests.
Our input map is shown in Figure 1.


\subsection{The simulation of the sky observation} \label{sec_30_10}

The instrument design for Planck mission
calls for multi-frequency focal plane arrays placed at the focus
of off-axis optical systems, in order to achieve proper angular resolution,
sensitivity, and spectral coverage. As a consequence, not all
the feedhorns can be located
very close to the centre
of the focal plane; the study of the implications of the related
beam optical distortions on temperature measurements has been presented in other
works (Burigana et al. 1997b,c).
For the present purpose  we
do not convolve the sky map with the beam response, but we simply read
the map temperature corresponding to the pixel identified by
the central direction of a given beam during the sky scanning.

Figure 2 shows the schematic representation of the observational
geometry.



Let $i$ be the angle between a unit vector $\vec s$, along the satellite
spin axis (outward the Sun direction),
and the normal to the ecliptic plane,
and $\vec p$ the unit vector of the direction of the optical axis of the telescope,
at an angle $\alpha$ from the spin axis ($i=90\deg$ and $\alpha=70\deg$ for
the Phase A study (Bersanelli et al. 1996).
We choose two coordinates $x$ and $y$ on the plane tangent to the celestial
sphere in the telescope optical axis direction,
with unit vector $\vec u$ and $\vec v$ respectively;
we choose the $x$ axis according
to the condition that the unit vector $\vec u$ points always toward the
satellite spin axis; indeed, for standard Planck observational strategy,
this condition is preserved as the telescope scans different sky regions.
With this choice of reference frame, we have that
$\vec v = \vec p \wedge \vec s / |\vec p \wedge \vec s|$ and
$\vec u = \vec v \wedge \vec p / |\vec v \wedge \vec p|$
(here $\wedge$ indicates the vector product).
In general the coordinates $(x_0,y_0)$ of the beam centre
will be identified by two angles; we use here the colatitude $\theta_B$ and
the longitude $\phi_B$
in the $\vec u , \vec v , \vec p$ reference frame
(see the Appendix A for details on geometrical
transformations).
For the present test we
assume a typical off-axis location of the considered
beam: $\theta_B=2.8^{\circ}$,
$\phi_B=45^{\circ}$. 
We note that our choice of $\theta_B$ is representative of typical
LFI beam position for a telescope with a primary mirror
of 1.5m aperture in the new Planck optical configuration
(see Mandolesi et al. 1997b). Our choice of $\phi_B$ 
corresponds to a case of intermediate efficiency respect 
to the destriping technique; the cases $\phi_B=0^{\circ}$ 
(beam located along the $\vec u$ direction in the TICRA U-V plane)
and $\phi_B=90^{\circ}$ 
(beam located along the $\vec v$ direction in the U-V plane)
are equivalent respectively to an on-axis beam and to
a beam that suitably distributes the 
crossings between different circles in two regions, close to the 
ecliptic poles, of maximum size.
The off-axis choice for $\theta_B$, when $\phi_B$ significantly
differs from 0, also assures  that,
even in the particular case of an angle of $90^{\circ}$ between
the spin axis and the telescope direction,
the scanning circles 
do not cross always exactly the ecliptic poles but
two somewhat larger regions around them also for 
scanning circles corresponding to significantly different 
spin axis positions.

%

We remember that 
in the Planck scanning strategy of the Phase A study
the sampling time of each receiver 
is choosen in order 
to have three samplings when the telescope axis describes
in the celestial sphere an angle with a length equal to the beam FWHM
($\sim$ three samplings per beam).
For any value of the angle $\alpha$, this condition determines
the number of samplings, $n_p$, per scan circle.

We will study in detail the effect introduced by the telescope motion
in a future work. For the present purpose, we 
simply assign the proper beam directions
to each sampling, extract the corresponding
pixel and read its temperature in the simulated input map; 
given the assumed FWHM of $30'$ and the pixel dimension
$19.4'$ we have typically 
2--3 different pixels explored in an integration time corresponding
to 3 samplings.

We remember that, in order to be able
to reduce the white noise in a simple way,  
we want to ``close'' the scanning circle, i.e. we need to modify
just a little 
the integration time by requiring that the telescope points always at
the same set of directions when it repeats the (120 for the Phase A study)
number of cycles with the same spin axis direction. 
Finally we have applied a further
reassessment of the integration time to make $n_p$ multiple of 12, 
in order to be able of applying in the future the destriping technique
not only with one but also with 2, 3 or 4 level constants for circle
(see sections~\ref{sec_50} and~\ref{sec_80}).

Our code let us free to implement arbitrary scanning strategies
(see also Appendix A);
we consider here 
cases with the spin axis always on the ecliptic plane and with an angle
$\alpha$ between the spin axis and the
telescope direction of $80^{\circ}, 85^{\circ}, 90^{\circ}$;
lower values of $\alpha$,
like that of $70^{\circ}$ assumed for the Phase A study, 
 will require wide oscillations 
($15^{\circ} - 20^{\circ}$)  
of the spin axis on the ecliptic plane 
(with relevant problems for the thermal stability)
in order to observe the regions close to ecliptic poles,
that are indeed very informative, being not significantly contaminated 
by the galactic emission. We have considered also a case with 
$\alpha = 90^{\circ}$ and with ten $10^{\circ}$ 
sinusoidal oscillations of the spin axis on the ecliptic plane.
For all the cases we have considered a 360 days mission
in order to compare results obtained with 
the same mission duration.

\subsection{The generation of instrumental noise series}

We generate white noise and 1/f noise using a random number generator
code (see Press et al. 1992).
For the white noise this is very simple: we have simply to rescale
a gaussian random noise distribution, by taking into account
the radiometer $rms$ white noise [see eq.~\ref{eq_10}].
For including the $1/f$-type noise we have adapted in FORTRAN 
an original IDL code
provided us by M. Seiffert, based on the power spectrum expansion
in the Fourier space of the noise components.
We generate together white and $1/f$ noise. 
Firstly a random gaussian distribution of (white) noises is generated;
then we calculate its power spectrum using an FFT code
(see Press et al. 1992); the amplitude of this power spectrum 
is then multiplied by $(1+f_k/f)^{1/2}$ for including 
the $1/f$ contribution.
Finally we have the time series with both the noises 
by computing the (inverse) FFT of the power spectrum.
When we take into account the real properties of the 30 GHz receivers, 
eq. (9) (Seiffert et al. 1997) gives
the $1/f$ knee frequency of our radiometers. 
We adopt as reference for the present simulations $f_k=0.05$Hz 
a value adequate to a cooling efficiency that allows to keep 
a load temperature $T_y \simeq 20$K (we will use here 
$T_n=9$K, $\beta = 6$ GHz).
Our noise series consist of about $2 \times 10^6$ evaluations and
a single series covers the integrations for 8 different contiguous
spin axis directions (16 hours).
We use a single series for white noise and $1/f$ noise together, but 
we generate also a noise series that includes white noise only;
this increases the computation time of only few percent (see above)
but it is useful for comparison and 
for the quantification of the stripes magnitude
and of the destriping efficiency 
(see sections $6 \div 10$).
In our code the generation of noise series is coupled with
the sky observation; more precisely, because of the noises dependence
on the antenna temperature $T_x$, we include the exact local sky
temperature in the noise magnitudes. 
This kind of flexibility may be useful also in the
future for simulations that will include Galaxy emission and
thermal drifts too.
Then, the only simplification is 
in the estimate of the knee frequency for the $1/f$ noise series, which
is assumed to be constant (i.e. with constant $T_x$, 
not allowing for spatial or time variations); 
the adopted value is consistent with the CMB monopole
antenna temperature and a typical environment temperature of 1 K. 

\subsection{The data stream generation and recording} \label{sec_30_30}

We have recorded our data streams in 4 matrices; any row of these matrices
refers to the data obtained from a given spin axis direction;
the number of columns is equal to the number of samplings
($n_p \sim 2100$, depending on $\alpha$);
the number of rows is equal to
the number, $n_s$,  of different spin axis directions
(4320 for a shift of 5' in the spin axis direction).
We have recorded the following data:
\begin{itemize}

\item the matrix $\bf N$ which contains the pixel numbers -- 
4 bytes per pixel --
  (in COBE-cube pixelisation) corresponding to the different integrations;

\item the matrix $\bf T$ which contains the "global" temperatures observed by the
  receiver in the above sky directions, directly averaged
  over the number of cycles per scan circle 
  (120 in the Phase A study scheme) in order to avoid
  the use of a large useless amount of memory space.
  $\bf T$ includes:
  the "input" sky temperature fluctuation as observed in the
  adopted geometrical scheme, which obviously can be computed a single time
  for any given spin axis direction;
  the white noise and the 1/f noise averaged over the number of cycles
  for each spin axis direction;

\item the matrix $\bf W$, generated
  in the same way as the matrix $\bf T$, but 
  containing the temperatures that will be observed in presence
  of the white noise only (see section 3.2);
  
\item the matrix $\bf G$ contains the temperatures that will be observed in 
  absence of instrumental noise and will be useful
  to check the goodness of the geometrical part of our flight simulation
  code.

\end{itemize}

Of course the matrices $\bf W$ and $\bf G$ do not have the corresponding
ones in a real observation.

In principle, if we simulate the mission for a time long enough that the
spin axis return on the same direction after a certain period 
(360 days, as a reference allowing for the case $\alpha = 90^{\circ}$) 
we can average the data of the second period
with the  corresponding rows of the first period, and take memory that
their will be affect by a (statistical) error 
$\sqrt 2$ times smaller than that corresponding to rows that are observed
for a single period. For a more realistic simulation,
we need to record a further matrix $\bf E$ with the statistical sensitivities
corresponding to the pixels of matrix $\bf N$,
which  takes into account this fact as well as a possible degradation in
sensitivity for some elements of data streams due for example to cosmic rays,
spurious effects ... . We neglect these kinds of complications for the present
analysis, which is equivalent to say that all the elements of $\bf E$ assume
a constant value. Nevertheless, in section~\ref{sec_50} we draw out our formulas
by including this possible effect, for sake of generality.

For implementing the destriping techniques (see section 5)
we need to recognize when the pointing directions for different spin axis
directions are substantially identical. 
The sky pointing direction is stored in this scheme only 
through the corresponding pixel number. 
Then, from a statistical point of view, two pointing directions 
are considered identical provided that their distance is 
smaller than the pixel size. 
The resulting number of pixels in common is then related to the assumed
pixel size. We have then recorded:

\begin{itemize}

\item additional matrices $\bf N_H$ (H=1,2,...,$n_R$) contains the pixel  number
corresponding to the different integrations for a certain number, $n_R$, of
resolutions higher than that used for the input/output maps
(for example at resolution 10 or 11 for input map at resolution 9).
By exploiting
these matrices we will able to test the adopted destriping technique 
under more stringent conditions on the average distance 
between their pointing directions
in the research of the pixels in common.

\end{itemize}

\section{From data streams to observed simulated maps} \label{sec_40}

Given the above simulated data streams, it is quite simple to obtain the following
simulated observed maps (see also section~\ref{sec_50_20}),
that can be easily compared one to each other.

We compute the sensitivity map, $\bf M_S$, 
with which any map pixel is observed, by recognizing
how many times a given pixel is observed from the analysis of
the whole matrix $\bf N$ (and $\bf E$ if this is the case).

From the matrices $\bf N$ and $\bf T$, we average the temperatures corresponding
to the same pixel in different matrix positions to have the
observed temperature map $\bf M_T$ (including noises).

In similar way, from the matrices $\bf N$ and $\bf W$ (or $\bf G$) 
we obtain the
observed temperature map $\bf M_W$ (or $\bf M_G$) 
computed in presence of white noise only
(or in absence of instrumental noises).
We have verified that
the map $\bf M_G$ is identical to the ``input'' map
for all the observed pixels, so confirming the 
validity of the geometrical part of our flight simulation code.

\section{Destriping techniques} \label{sec_50}

We have developed a technique in order to eliminate the effects of gain
drifts in Planck signal due to the 1/$f$ noise effect. The method is
derived from that proposed for the Phase A Study and re-analyzed
by Delabrouille (1997). On the other hand, our treatment of
the Planck observation simulation, although simplified,
is general enough to be close to the ``real'' Planck observations; 
so we draw out here below the destriping
mathematical formalism, in a way directly applicable to our
simulated data streams.

\subsection{Mathematical formalism} \label{sec_50_10}

In this section we discuss how to eliminate the effects of gain
drifts on timescales greater than that for which the spin axis points
at a given direction (2 hours for the Phase A scheme,
or 1 hour or 40 minutes for possible new observational strategies
in which the spin axis shift of 2.5' or 1.6' in 1 hour or 40 minutes respectively),
i.e. the satellite scans a given circle in the sky.

After we removed the drifts within any given scan circle
by averaging the observations over the corresponding cycles
(see section 3.1),
each set of observations at a given spin axis direction, denoted by
the index $i$, is characterized
by an additive level $A_i$ which is related to the ``mean''
$1/f$ noise level during the observation in that scan circle.
These levels $A_i$ are different for different circles, due to
gain fluctuations. Our goal is to obtain a reduced set of observations
of different scan circles by removing the contamination
that affects any circle. So we will subtract to all sets of observations
on a given scan circle their own characteristic level $A_i$.
As a variance, we can attribute more constant levels, say $n_l$ levels, 
per single scan circle. From the computational point of view
this is exactly equivalent to rearrange all the matrices 
in section 3.3 by dividing
their rows in $n_l$ parts that have to be appropriately relocated 
to construct new matrices with $n_s \times n_l$ rows and
$n_p / n_l$ columns that can be then analysed exactly as in the case
of a single constant per circle. 

For estimating all these levels we use a computation scheme able to
simultaneously find the pixels in common between different scan circles
and generate a linear system whose solution gives the
unknowns $A_i$.

The observations of different directions in the sky explored by the
satellite have been recorded in three matrices ${\bf N}$, ${\bf T}$, ${\bf E}$
of $n_s$ rows
and $n_p$ columns, where $n_s$ is the number of different spin axis directions,
and $n_p$ is the number of samplings at different horn pointing directions
in a given circle.
Here $N_{il}$ ($i=1, \dots, n_s$, $l=1, \dots, n_p$) contains
the pixel number corresponding to the observed direction in the sky,
and $T_{il}$ and $E_{il}$ are respectively the corresponding observed
temperature (full signal due to the sky plus noises)
and the estimates of the $rms$ noise, essentially due to the white noise. 
We observe that $E_{il}$ is properly related to the amplifier noise temperature
and to the observed antenna temperature which depends 
in a real case on the local thermal
conditions and on the true sky temperature but it may depend also
on possible ``spurious effects''. We have in reality only ``first-order''
informations about all these quantities at this level;
on the other hand its accurate knowledge is not crucial, being the amplifier
noise temperature typically higher than the
observed temperature.
Anyway we hope to have more accurate informations from accurate thermal models
and from iterating our data reduction scheme to have a good determination of
sky temperature, which is of course our goal.

In the following the first index will denote the
row index and the second the column index.

We must check for all the possible crossing points for
two different circles for the whole ensemble of $n_s$ circles (see also
section 3.3).

Let $\pi$ be an index that identifies a generic couple of different
observations corresponding to the same pixel in the sky, i.e. a pixel
in common between two scan circles: $\pi$ ranges from $1$ to $n_c$, where
$n_c$ is the total number of couples found.
Therefore the index $\pi$ is related to two elements in the matrix ${\bf N}$:
$\pi \to (il, jm)$. Here $i$ and $j$ identify the two circles for which 
we found common pixels, $l$ and $m$ are the common pixels positions on 
the circle $i$ and on the circle $j$ respectively. So we have $N_{il}=N_{jm}$.
As a variance, we can replace the matrix $\bf N$ with one of the matrices
$\bf N_H$ (see section 3.3 and 7.4) to search for the pixels in common,
according to the adopted averaged maximum distance for recognizing two pixels
in common.

We want to minimize the quantity:
\begin{equation}
  S = \sum_\mathrm{all\; couples} 
  \left[ \frac{[(A_i - A_j) - (T_{il} - T_{jm})]^2}
    {E_{il}^2 + E_{jm}^2} 
  \right] = \sum_{\pi=1}^{n_c} 
  \left[ \frac {[(A_i - A_j) - (T_{il} - T_{jm})]^2}
    {E_{il}^2 + E_{jm}^2} 
  \right]_\pi
\end{equation}
respect to the set of the unknown levels $A_i$; the index $\pi$ in the right 
hand side of this equation remembers that each set $(il,jm)$ derive from a
given pixel $\pi$.
S is quadratic in all the unknown $A_i$;
on the other hand, only the differences between the levels $A_i$
enter in this expression, so that the solution will be
indeterminate, i.e. the levels are determined apart from an
arbitrary additional constant (with no physical meaning, as obvious for
anisotropy measurements).

To remove this indetermination, we add a constraint to the $A_i$ quantities:
\begin{equation} \label{eq_100}
  \sum_{h=1}^{n_s} A_h = 0 \, .
\end{equation}
This is equivalent to minimize the quantity:
\begin{equation}
  S' = S + \left( \sum_{h=1}^{n_s} A_h \right)^2
\end{equation}

Now let's go into some algebra. We perform the derivate of the previous
equation, and finally we have:
\begin{equation} \label{eq_110}
  \frac 12 \frac{\partial S'}{\partial A_k} = \sum_{\pi=1}^{n_c} 
  \left[ \frac {[ (A_i - A_j) - (T_{il} - T_{jm}) ] \cdot
    \left[ \delta_{ik} - \delta_{jk} \right]} {E_{il}^2 + E_{jm}^2}
  \right]_\pi + \sum_{h=1}^{n_s} A_h = 0
\end{equation}
for \emph{all} $k = 1, \dots, n_s$ (here the $\delta$ are the usual Kronecker
symbols). So we have a set of $n_s$ equations:
\begin{equation} \label{eq_120}
  \sum_{t=1}^{n_s} C_{kt} \ A_t = B_k \, , \quad k=1, \dots, n_s \, .
\end{equation}
We denote with ${\bf C}$ and ${\bf B}$, respectively, the matrix of
the coefficients $C_{kt}$ and the vector of the coefficients $B_k$.

To be concrete, we show here as ${\bf C}$ and ${\bf B}$ are formed
as we extract the pixels in common between the different rows.
First of all, we set $\mathbf{B} = 0$ and $C_{kt} = 1$~$\forall k,t$ (setting
all $C_{kt}$ to $1$ takes into account the second term of eq.~\ref{eq_110}).
Then for each couple $\pi$ of pixels in common between two scan circle we
define: 

\begin{equation}
  \chi_\pi=\left[ \frac 1 {E_{il}^2 + E_{jm}^2} \right]_\pi
\end{equation}
and
\begin{equation}
  \tau_\pi=\left[ \frac {T_{il} - T_{jm}} {E_{il}^2 + E_{jm}^2} \right]_\pi
\end{equation}

From the above equation and the definition of Kronecker symbol, we easily have
that a given couple $\pi$ contributes only to two equations of our
linear system, those for $k=i$ or $k=j$, where as usual $i$ and $j$
corresponds to two different observations of the same pixel.
If we iteratively increment the coefficients
of ${\bf C}$ and ${\bf B}$ as we find a new couple, explicitely we
have [remember $\pi \to (il, jm)$]:
\begin{gather}
  C_{ii} \to C_{ii} + \chi_\pi \\
  C_{ij} \to C_{ij} - \chi_\pi \\
  C_{ji} \to C_{ji} - \chi_\pi \\
  C_{jj} \to C_{jj} + \chi_\pi \\
  B_i \to B_i + \tau_\pi \\
  B_j \to B_j - \tau_\pi
\end{gather}

Summing up, we have that each couple $\pi$ contributes to only six terms,
and the resulting system shows a complete symmetry with respect to the 
exchange of the indexes $i$ and $j$. 

The linear system defined by eq. (\ref{eq_120}) has some interesting
properties which considerably simplify the numerical computation of 
its solution. In particular, the matrix $\bf C$:
\begin{itemize}
\item is \emph{symmetric}, so we can hold in memory only half of the matrix
(say the upper-right part) and solve the system and speed-up the code by 
computing only half of the matrix coefficients. This is possible because
the Gauss reduction algorithm preserves at each step the symmetry of
remaining part of the matrix;

\item is \emph{positive defined}, so we never find a null pivot when
reducing a non-singular matrix (Strang 1976). This allow us to solve
the system without having to exchange rows or columns, so preserving the
symmetry;

\item is \emph{not singular} (provided that there are enough intersections
between scan circles), because the only fundamental indetermination has
been removed by imposing the constraint (\ref{eq_100}).

\end{itemize}  

Anyway, after the system has been solved, it is our care to replace the
solution into the original $\bf C$ matrix, to verify its correctness and
check for rounding errors and/or accidental degenerations.

\subsection{Some remarks on numerical efficiency, RAM requirements and
off-sets} \label{sec_50_20}

In order to speed the construction of the matrix $\bf C$ and of the vector
$\bf B$, we have found that it is very advantageous
to firstly order (we use the quick sort algorithm)
all the elements of the matrix $\bf N$, i.e. the 
observed pixels (4 bytes integers), 
in the first column of a new ``matrix'' $\bf U$ (of $n_p \times n_s$ rows 
and 3 ``columns'') by keeping memory of 
their locations (2$\times$2 bytes integers), in the original matrix $\bf N$
in the other two ``columns'' of $\bf U$.
In this way we simply extract once for all each pixel in common
between two scan circles for all scan circles, being the same pixel
located in contiguous rows in the matrix $\bf U$, by considering all the 
possible pairs of rows of $\bf U$ with the same element in the first column, 
with the simple caution that the elements of the second column of $\bf U$, 
i.e. the original rows in the matrix $\bf N$, are different.
In this way the ``scanning'' of the matrix $\bf N$ and the construction
of $\bf C$ and $\bf B$ according to the rules of section~\ref{sec_50_10} turns
to be very fast. It is immediate to use the matrices $\bf N_H$, 
containing the pixel numbers at higher resolutions, in the construction of 
$\bf C$ and $\bf B$ if one want to adopt more stringent conditions
on the distance between pixels in common.
In addition, working with the additional matrix $\bf U$ optimize
the construction of the simulated maps from the simulated
data streams (see section~\ref{sec_40}), being immediate to recognize in the
matrix $\bf U$ when the same pixel has been observed.

For the solution of the linear system (\ref{eq_120}) 
we have found that the Gauss elimination method works very well.
We prefer to construct and solve the system by using double
precision accuracy, to have high numerical accuracy and to be sure
of avoiding artificial numerical singularity; also, due to 
matrix symmetry and positive definiteness, we do not need pivot.

To build up 
the linear system, we have to keep the memory space 
for the system matrix, the system known terms, the observed temperature
matrix and the auxiliary (integer) matrix $\bf U$; 
by taking advantage of this symmetry, and by considering in general
$n_l$ constant levels per scan circle the memory requirement
is: $8 {\rm bytes} \times [n_l n_s (n_l n_s + 1)/2 + n_l n_s + n_s n_p]
+ 4 {\rm bytes} \times 2 \times n_s n_p$. 
For example, at 30 GHz (FHWM$\simeq 30'$, $n_p \simeq 2100$), 
for the case of $5'$ shift of the spin axis ($n_s=4320$)
we need about 220 (440) Mbytes by working with $n_l=1$ ($n_l=2$);
for a $2.5'$ shift of the spin axis, $n_s=8640$ and we need
about 590 (1500) Mbytes by working with $n_l=1$ ($n_l=2$).
For sake of illustration, if we have a beam of $\simeq 10'$
(like the nominal 100 GHz beams) and we want to
record 4 samplings per beam we will have $n_p \simeq 8400$;
for a $2.5'$ shift of the spin axis and by working with
$n_l=1$ ($n_l=2$) the memory requirement is of about 1500 (2400) Mbytes.
This memory problem can be solved by taking advantage of disk buffers; 
we discuss our solution in the Appendix B.

For solving the system we only need to keep in memory
the system matrix and known terms, and 
the memory requirement
is: $8 {\rm bytes} \times [n_l n_s (n_l n_s + 1)/2 + n_l n_s]$. At 30 GHz
and with 3 samplings per beam, we need about 
75 Mbytes if $n_s=4320$ and $n_l=1$, 300 Mbytes
if $n_s=4320$ and $n_l=2$ or $n_s=8640$ and $n_l=1$ and
1200 Mbytes if $n_s=8640$ and $n_l=2$. This problem 
maybe crucial depending on the available amount of RAM, especially
because the Gauss elimination continuously changes
the system components.
Also this memory problem can be solved by taking advantage of disk buffers 
(Appendix B).


After the solution of the linear system, 
we obtain
a ``destripped'' matrix $\bf D$
by subtracting the level $A_i$ to
the $i-$th row ($i=1,n_s$) of the  matrix $\bf T$ 
Then we apply to this matrix the same treatment of section~\ref{sec_40}
and we have the observed destripped temperature map $\bf M_D$.

We observe that,
contrary to the case of pure white noise,
the average of a $1/f$-type noise series can be 
significantly different from zero.
Then, the map $\bf M_T$ (as well as the map $\bf M_D$, but in general
with a somewhat different value)
may present an off-set with respect to the map $\bf M_G$; on the contrary
the off-set between the map $\bf M_W$ and the map $\bf M_G$ is negligible.
As a typical example, for the simulation with $\alpha =90^{\circ}$ 
(see Table 1) we find
that these off-sets are $\simeq 4.8 \mu$K; for comparison,
the off-set we find between the maps $\bf M_W$ and $\bf M_G$ is much smaller,
$\simeq 0.075 \mu$K.
The off-set between $\bf M_T$ ($\bf M_D$, $\bf M_W$) and $\bf M_G$
is of course not relevant for anisotropy measurements,
nor we are able to subtract it in a real case
(on the other hand we must pay attention to the fact that off-sets
may be present between maps produced by different receivers at the same
frequency).
The off-set of the map $\bf M_T$ ($\bf M_D$, $\bf M_W$) 
must be removed by subtracting
the difference between the average of the map $\bf M_T$ 
($\bf M_D$, $\bf M_W$) and of the
map $\bf M_G$: this is necessary for a correct quantitative 
analysis of stripes magnitude and destriping efficiency.

We indicate with $\bf {\widetilde M_T}$, $\bf {\widetilde M_D}$
(and $\bf {\widetilde M_W}$, but it is not relevant in practice for this matrix)
the above maps, when this kind of off-set has been removed.

\section{Estimators of the destriping efficiency} \label{sec_60}

In the previous sections we have described our
simulations of the Planck observations and the basic treatment
to convert observational data streams into sky maps,
including destriping techniques.
Here we analyse the efficiency of the adopted destriping technique,
by considering well known estimators.

\subsection{Ratio between the $\chi^2_r$'s} \label{sec_60_10}

We expect that the average of the squares
of differences between the elements of the maps
$\bf {\widetilde M_W}$ and $\bf M_G$ divided by the observation sensitivity
[essentially the estimator $\chi^2_r$; we will call it
in this case $(\chi^2_r)_W$]
is very close to 1, because only the white noise is present
in this case.
We can compute the same estimator for  $\bf {\widetilde M_T} - M_G$
[$(\chi^2_r)_T$, undestripped case]
and  $\bf {\widetilde M_D} - M_G$ [$(\chi^2_r)_D$ destripped case].
(For the above consideration,
by using observed maps without removing the off-sets
one will find a meaningless amplification of the $\chi^2_r$).

On the other hand, we find that the exact value of
$(\chi^2_r)_W$ may be just a little different from 1 depending indeed 
on the assumed sensitivity; for example it is just a little 
different is we divide $\bf {\widetilde M_W} - M_G$ by the map $\bf M_s$
(i.e. by using the sensitivity proper of any pixel -- we will use this definition
in our tables) or the average
of the sensitivities in the map $\bf M_S$ or the estimate of the
average sensitivity
obtained on the basis the global mission time, the observed
number of pixels (393216 for our maps
at COBE-cube resolution 9) and the properties of considered receiver. 
Then we prefer to use the
ratio between the $(\chi^2_r)_T$ -- or $(\chi^2_r)_D$ --
and $(\chi^2_r)_W$ as estimator.
This ``renormalized'' estimator
(that we will denote by $\chi^2_{r,n,T}$ and  $\chi^2_{r,n,D}$
respectively for the undestripped and destripped cases)
results independent
of the choice of the sensitivity adopted for
the $\chi^2_r$ calculation, so allowing a better understanding of
the magnitude of striping effect and of destriping efficiency.
We will quantify the destriping efficiency by using the relative decrease
of the renormalized $\chi^2_r$, i.e. with the quantity
$[(\chi^2_r)_D - (\chi^2_r)_T]/[(\chi^2_r)_T -1]$. 

\subsection{Magnitude of stripes temperature}

From the values of $\chi^2_{r,n,T}$ and $\chi^2_{r,n,D}$ defined above
and from the average $rms$ white noise, $rms_W$, for the observed pixels 
derived from the sensitivity map $\bf M_S$
we can easily give an estimate of the $rms$ temperature 
of the stripes before, $rms_T$, and after destriping, $rms_D$. 
Under the hypothesis that the error introduced by the noises
in each pixel may be thought as a sum of two uncorrelated contributions from
white noise and $1/f$ noise, we have  
$rms_T = rms_W \sqrt{\chi^2_{r,n,T} -1}$ and 
$rms_D = rms_W \sqrt{\chi^2_{r,n,D} -1}$ [Method (a)]. 


Another estimate [Method (b)] of the $rms$ stripes temperature can be obtained
by directly evaluating the global temperature $rms$ difference
before, $rms_{tot,T}$, or after, $rms_{tot,D}$, the destriping from the
comparison with the maps $\bf M_G$ and by assuming that they are given
by the sum in quadrature of $rms_W$ and $rms_T$ or $rms_D$. 
From the values of $rms_W$, $rms_{tot,T}$ and $rms_{tot,D}$ we can calculate
$rms_T$ or $rms_D$. 

Finally [Method (c)], we can treat the $1/f$ contribution to the total noise
like a systematic (and not statistical) error and therefore to 
assume that $rms_{tot,T}$ or $rms_{tot,D}$ are simply given
by the sum of $rms_W$ and $rms_T$ or $rms_D$. 

For estimating the destriping efficiency in terms of residual stripes 
temperature we will by using the relative decrease
of the stripes temperature, i.e. the quantity
$(rms_D - rms_T)/rms_T$.

\section{Results} \label{sec_70}

From the visual inspection of the simulated maps the effect of the noises is of
course not clearly evident; we only recognize a somewhat degradation
of the map details. The stripes become more evident when we plot
the noises map only (see Figure 3); 
they can be obtained by subtracting the 
CMB fluctuation map ($\bf M_G$) to the observed map.
The stripes figures well reproduce the adopted scanning strategy.


\subsection{Destriping versus scanning strategy}

The visual inspection of our ``stripes'' maps 
do not allow to quantify the
striping magnitude and its reduction obtained from the destriping procedure
(see Figures 3 and 4).


The statistical analysis of the maps allows a much better understanding
of the destriping procedure efficiency. 
We present here (see Tables $1 \div 4$) the results 
of our simulations (for a typical channel at 30 GHz)
in terms of reduced $\chi^2_{r,n}$ 
and of stripes temperature for the three choosen 
values of the (constant) angle $\alpha$ between
the spin axis and the telescope direction and for the considered case
with $\alpha = 90^{\circ}$ and spin axis oscillations
(see also the Appendix A).
We report also in the tables some informations on relevant quantities:
the ideal white noise for a single sampling time which weakly increases with
$\alpha$ for geometrical reasons if we want to have 3 samplings per beam;
the (single receiver) average $rms$ white noise, $rms_W$, 
(expressed in mK) for the observed pixels for a 360
days mission; 
the square of the ratio $R$ between
these white noises, normalized
to the intermediate case of $\alpha = 85^{\circ}$;
the percentage of sky which results to be observed by
the considered single off-axis beam, which also increases with $\alpha$;
the number of couples in common find in the destriping procedure and
the number of constant levels per scan circle used in the destriping
code;
the map resolution and the resolution used for searching the pixels in common.
We remember the adopted value, 0.05 Hz, 
of $1/f$ knee frequency and of the bandwidth, 6 GHz,
and that the sampling time is of about 
0.03 sec, the exact value depending on the choosen value of $\alpha$.  
In the tables we report our values of $\chi^2_{r,n}$
before and after the destriping procedure and the relative (\%) decrease 
without the multiplicative factor $R^2$ and by taking it into account.
Indeed the (white noise) sensitivity per pixel is different for different scanning
strategies; then, by including the factor $R^2$ we ``renormalize''
the values of $\chi^2_{r,n}$ at the same (white noise) sensitivity level, 
so making the them essentially independent of the $\alpha$--dependent sensitivity.

\bigskip
\bigskip
\begin{center}
  \begin{tabular}{cccc}
    \multicolumn{4}{c}{{\bf Table 1}: destriping results; $\alpha=90^{\circ}$.} \\
    \hline
    \hline
    & & &\\
    Some global parameters\\
    \hline
    & & & \\
$\Delta T_W = 1.322 $mK  &  $rms_W = 27.18 \mu$K & $R^2 = 1.00666 $ & \% sky = 99.98 \\
Map Res. = 9 & Res. common pix. = 9 & $n_l = 1 $ & pix. in common = $2.70 \times 10^8$ \\
    & & & \\
    \hline
    & & & \\
Before destriping & After destriping & \% improvement & Method \\
    \hline
    & & & \\
$\chi^2_{r,n,T} = 1.1770$  &  $\chi^2_{r,n,D} = 1.0216$   &  87.8 \% & \\  
$\chi^2_{r,n,T} R^2 = 1.1848 $  &  $\chi^2_{r,n,D} R^2 = 1.0284 $   & 84.6 \% & \\  
$rms_T = 11.4 \mu$K  &  $rms_D = 3.99 \mu$K   &  65.1 \% & (a) \\  
$rms_T = 17.3 \mu$K  &  $rms_D = 6.93 \mu$K   &  59.9 \% & (b) \\  
$rms_T = 5.03 \mu$K  &  $rms_D = 0.87 \mu$K   &  82.7 \% & (c) \\  
    & & & \\
    \hline
    \hline
  \end{tabular}
\end{center}


%
%


\bigskip
\bigskip
\begin{center}
  \begin{tabular}{cccc}
    \multicolumn{4}{c}{{\bf Table 2}: destriping results; $\alpha=85^{\circ}$.} \\
    \hline
    \hline
    & & &\\
    Some global parameters\\
    \hline
    & & & \\
$\Delta T_W = 1.318 $mK  &  $rms_W = 27.09 \mu$K & $R^2 = 1$ & \% sky = 99.43 \\
Map Res. = 9 & Res. common pix. = 9 & $n_l = 1 $ & pix. in common = $2.34 \times 10^8$ \\
    & & & \\
    \hline
    & & & \\
Before destriping & After destriping & \% improvement & Method \\
    \hline
    & & & \\
$\chi^2_{r,n,T} = 1.2709 $  &  $\chi^2_{r,n,D} = 1.0142 $   & 94.7 \% & \\  
$\chi^2_{r,n,T} R^2 = 1.2709 $  &  $\chi^2_{r,n,D} R^2 = 1.0142 $   & 94.7 \% & \\  
$rms_T = 14.1 \mu$K  &  $rms_D = 3.23 \mu$K   & 77.1 \% & (a) \\  
$rms_T = 21.6 \mu$K  &  $rms_D = 7.31 \mu$K   & 66.2 \% & (b) \\  
$rms_T = 7.98 \mu$K  &  $rms_D = 0.969 \mu$K  & 87.9 \% & (c) \\  
    & & & \\
    \hline
    \hline
  \end{tabular}
\end{center}


%
%

\bigskip
\bigskip
\begin{center}
  \begin{tabular}{cccc}
    \multicolumn{4}{c}{{\bf Table 3}: destriping results; $\alpha=80^{\circ}$.} \\
    \hline
    \hline
    & & &\\
    Some global parameters\\
    \hline
    & & & \\
$\Delta T_W = 1.311 $mK  &  $rms_W = 26.90 \mu$K & $R^2 = 0.98602 $ & \% sky = 98.12 \\
Map Res. = 9 & Res. common pix. = 9 & $n_l = 1 $ & pix. in common = $2.20 \times 10^8$ \\
    & & & \\
    \hline
    & & & \\
Before destriping & After destriping & \% improvement & Method \\
    \hline
    & & & \\
$\chi^2_{r,n,T} = 1.0396 $  &  $\chi^2_{r,n,D} = 1.0292 $   &  26.19 \% & \\  
$\chi^2_{r,n,T} R^2 = 1.0251 $  &  $\chi^2_{r,n,D} R^2 = 1.0148 $   & 41.0 \% & \\  
$rms_T = 5.35 \mu$K  &  $rms_D = 4.60 \mu$K   &  14.1 \% & (a) \\  
$rms_T = 8.05 \mu$K  &  $rms_D = 6.94 \mu$K   &  13.8 \% & (b) \\  
$rms_T = 1.17 \mu$K  &  $rms_D = 0.88 \mu$K   &  24.8 \% & (c) \\  
    & & & \\
    \hline
    \hline
  \end{tabular}
\end{center}


%
%

\bigskip
\bigskip
\begin{center}
  \begin{tabular}{cccc}
    \multicolumn{4}{c}{{\bf Table 4}: destriping results; 
$\alpha=90^{\circ} \pm 10^{\circ}$ (10 sinusoidal oscillations).} \\
    \hline
    \hline
    & & &\\
    Some global parameters\\
    \hline
    & & & \\
$\Delta T_W = 1.322 $mK  &  $rms_W = 29.07 \mu$K & $R^2 = 1.15152 $ & \% sky = 100 \\
Map Res. = 9 & Res. common pix. = 9 & $n_l = 1 $ & pix. in common = $2.36 \times 10^8$ \\
    & & & \\
    \hline
    & & & \\
Before destriping & After destriping & \% improvement & Method \\
    \hline
    & & & \\
$\chi^2_{r,n,T} = 1.1165$  &  $\chi^2_{r,n,D} = 1.0153$   &  86.9 \% & \\  
$\chi^2_{r,n,T} R^2 = 1.2857 $  &  $\chi^2_{r,n,D} R^2 = 1.1691 $   &  40.8 \% & \\  
$rms_T = 9.92 \mu$K  &  $rms_D = 3.60 \mu$K   &  63.8 \% & (a) \\  
$rms_T = 15.2 \mu$K  &  $rms_D = 8.62 \mu$K   &  43.3 \% & (b) \\  
$rms_T = 3.73 \mu$K  &  $rms_D = 1.25 \mu$K   &  66.5 \% & (c) \\  
    & & & \\
    \hline
    \hline
  \end{tabular}
\end{center}

\bigskip
\bigskip


\subsection{Destriping versus $1/f$ knee frequency}

For the interesting case $\alpha = 90^{\circ}$ with no oscillations, we
carried out other 
a simulation with 
a much larger value 
of the $1/f$ knee frequency, 
$f_k=10$Hz,
of order of that expected for total power 
radiometers.



It is interesting to study the stripes effect
and destriping performance under 
this very pessimistic condition.

Tables 5 shows our results that 
have to be compared with those of Table 1,
based on the theoretical estimate of $f_k$ of our kind of 
radiometers.

\bigskip
\bigskip
\begin{center}
  \begin{tabular}{cccc}
    \multicolumn{4}{c}{{\bf Table 5}: destriping results; $\alpha=90^{\circ}$; $f_k=10$Hz.} \\
    \hline
    \hline
    & & &\\
    Some global parameters\\
    \hline
    & & & \\
$\Delta T_W = 1.322 $mK  &  $rms_W = 27.18 \mu$K & $R^2 = 1.00666 $ & \% sky = 99.98 \\
Map Res. = 9 & Res. common pix. = 9 & $n_l = 1 $ & pix. in common = $2.70 \times 10^8$ \\
    & & & \\
    \hline
    & & & \\
Before destriping & After destriping & \% improvement & Method \\
    \hline
    & & & \\
$\chi^2_{r,n,T} = 9.2846$  &  $\chi^2_{r,n,D} = 2.6769$   &  79.8 \% & \\  
$\chi^2_{r,n,T} R^2 = 9.3464 $  &  $\chi^2_{r,n,D} R^2 = 2.6947 $   & 79.7 \% & \\  
$rms_T = 78.2 \mu$K  &  $rms_D = 35.4 \mu$K  &  54.8 \% & (a) \\  
$rms_T = 258 \mu$K  &  $rms_D = 67.7 \mu$K   &  73.7 \% & (b) \\  
$rms_T = 233 \mu$K  &  $rms_D = 45.8 \mu$K   &  80.3 \% & (c) \\  
    & & & \\
    \hline
    \hline
  \end{tabular}
\end{center}

\bigskip
\bigskip

\subsection{Destriping with more than one constant per scan circle} \label{sec_80}

For the reference case $\alpha = 90^{\circ}$, both with
the theoretical prediction for $f_k$ and for the case with $f_k$ 
representative of total power radiometers,
we have applied our destriping code by using two constants per scan circle.
The results are shown in Tables 6 and 7 that must be compared  with
Tables 1 and 5 respectively.
We find that the use of more constant per circle does not help
the destriping technique.

\bigskip
\bigskip
\begin{center}
  \begin{tabular}{cccc}
    \multicolumn{4}{c}{{\bf Table 6}: destriping results; $\alpha=90^{\circ}$.} \\
    \hline
    \hline
    & & &\\
    Some global parameters\\
    \hline
    & & & \\
$\Delta T_W = 1.322 $mK  &  $rms_W = 27.18 \mu$K & $R^2 =  1.00666 $ & \% sky = 99.98 \\
Map Res. = 9 & Res. common pix. = 9 & $n_l = 2 $ & pix. in common = $2.70 \times 10^8$ \\
    & & & \\
    \hline
    & & & \\
Before destriping & After destriping & \% improvement & Method \\
    \hline
    & & & \\
$\chi^2_{r,n,T} = 1.1770$  &  $\chi^2_{r,n,D} = 1.0280$   &  84.2 \% & \\  
$\chi^2_{r,n,T} R^2 = 1.1848 $  &  $\chi^2_{r,n,D} R^2 = 1.0348 $   &  81.2 \% & \\  
$rms_T = 11.4 \mu$K  &  $rms_D = 4.55 \mu$K   &  60.1 \% & (a) \\  
$rms_T = 17.3 \mu$K  &  $rms_D = 7.62 \mu$K   &  56.0 \% & (b) \\  
$rms_T = 5.03 \mu$K  &  $rms_D = 1.05 \mu$K   &  79.2 \% & (c) \\  
    & & & \\
    \hline
    \hline
  \end{tabular}
\end{center}


%
%

%
%

\bigskip
\bigskip
\begin{center}
  \begin{tabular}{cccc}
    \multicolumn{4}{c}{{\bf Table 7}: destriping results; $\alpha=90^{\circ}$; $f_k=10$Hz.} \\
    \hline
    \hline
    & & &\\
    Some global parameters\\
    \hline
    & & & \\
$\Delta T_W = 1.322 $mK  &  $rms_W = 27.18 \mu$K & $R^2 =  1.00666 $ & \% sky = 99.98 \\
Map Res. = 9 & Res. common pix. = 9 & $n_l = 2 $ & pix. in common = $2.70 \times 10^8$ \\
    & & & \\
    \hline
    & & & \\
Before destriping & After destriping & \% improvement & Method \\
    \hline
    & & & \\
$\chi^2_{r,n,T} = 9.2846$  &  $\chi^2_{r,n,D} = 2.7584$   &  78.8 \% & \\  
$\chi^2_{r,n,T} R^2 = 9.3464 $  &  $\chi^2_{r,n,D} R^2 = 2.7768 $   & 78.7 \% & \\  
$rms_T = 78.2 \mu$K  &  $rms_D = 36.2 \mu$K   &  53.7 \% & (a) \\  
$rms_T = 258 \mu$K  &  $rms_D = 70.0 \mu$K   &  72.9 \% & (b) \\  
$rms_T = 233 \mu$K  &  $rms_D = 47.9 \mu$K   &  79.4 \% & (c) \\  
    & & & \\
    \hline
    \hline
  \end{tabular}
\end{center}

\bigskip
\bigskip

%
%



\subsection{Destriping versus distance conditions} \label{sec_87}

For the reference case $\alpha = 90^{\circ}$, both with
the theoretical prediction for $f_k$ and for the case with $f_k$ 
representative of total power radiometers,
we have applied our destriping code by using 
the map pixels at higher resolution to search for pixels
in common.
The results are shown in Tables 8 and 9 that must be compared  with
Tables 1 and 5 respectively.
We conclude that the use of a more stringent condition
to find the coincidences of the pointing directions in different
scan circles does not help the destriping technique.

\bigskip
\bigskip
\begin{center}
  \begin{tabular}{cccc}
    \multicolumn{4}{c}{{\bf Table 8}: destriping results; $\alpha=90^{\circ}$.} \\
    \hline
    \hline
    & & &\\
    Some global parameters\\
    \hline
    & & & \\
$\Delta T_W = 1.322 $mK  &  $rms_W = 27.18 \mu$K & $R^2 =  1.00666 $ & \% sky = 99.98 \\
Map Res. = 9 & Res. common pix. = 10 & $n_l = 1 $ & pix. in common = $6.97 \times 10^7$ \\
    & & & \\
    \hline
    & & & \\
Before destriping & After destriping & \% improvement & Method \\
    \hline
    & & & \\
$\chi^2_{r,n,T} = 1.1770$  &  $\chi^2_{r,n,D} = 1.0235$   &  86.7 \% & \\  
$\chi^2_{r,n,T} R^2 = 1.1848 $  &  $\chi^2_{r,n,D} R^2 = 1.0303 $   &  83.6 \% & \\  
$rms_T = 11.4 \mu$K  &  $rms_D = 4.73 \mu$K   &  58.5 \% & (a) \\  
$rms_T = 17.3 \mu$K  &  $rms_D = 7.16 \mu$K   &  58.6 \% & (b) \\  
$rms_T = 5.03 \mu$K  &  $rms_D = 0.928 \mu$K  &  81.6 \% & (c) \\  
    & & & \\
    \hline
    \hline
  \end{tabular}
\end{center}


%
%

%
%

\bigskip
\bigskip
\begin{center}
  \begin{tabular}{cccc}
    \multicolumn{4}{c}{{\bf Table $8'$}: destriping results; $\alpha=90^{\circ}$.} \\
    \hline
    \hline
    & & &\\
    Some global parameters\\
    \hline
    & & & \\
$\Delta T_W = 1.322 $mK  &  $rms_W = 27.18 \mu$K & $R^2 =  1.00666 $ & \% sky = 99.98 \\
Map Res. = 9 & Res. common pix. = 11 & $n_l = 1 $ & pix. in common = $1.62 \times 10^7$ \\
    & & & \\
    \hline
    & & & \\
Before destriping & After destriping & \% improvement & Method \\
    \hline
    & & & \\
$\chi^2_{r,n,T} = 1.1770$  &  $\chi^2_{r,n,D} = 1.0334$   &  81.1 \% & \\  
$\chi^2_{r,n,T} R^2 = 1.1848 $  &  $\chi^2_{r,n,D} R^2 = 1.0403 $   &  78.2 \% & \\  
$rms_T = 11.4 \mu$K  &  $rms_D = 5.46 \mu$K   &  52.1 \% & (a) \\  
$rms_T = 17.3 \mu$K  &  $rms_D = 8.10 \mu$K   &  53.2 \% & (b) \\  
$rms_T = 5.03 \mu$K  &  $rms_D = 1.18 \mu$K   &  76.5 \% & (c) \\  
    & & & \\
    \hline
    \hline
  \end{tabular}
\end{center}


%
%

%
%

\bigskip
\bigskip
\begin{center}
  \begin{tabular}{cccc}
    \multicolumn{4}{c}{{\bf Table 9}: destriping results; $\alpha=90^{\circ}$; $f_k=10$Hz.} \\
    \hline
    \hline
    & & &\\
    Some global parameters\\
    \hline
    & & & \\
$\Delta T_W = 1.322 $mK  &  $rms_W = 27.18 \mu$K & $R^2 =  1.00666 $ & \% sky = 99.98 \\
Map Res. = 9 & Res. common pix. = 10 & $n_l = 1 $ & pix. in common = $6.97 \times 10^7$ \\
    & & & \\
    \hline
    & & & \\
Before destriping & After destriping & \% improvement & Method \\
    \hline
    & & & \\
$\chi^2_{r,n,T} = 9.2846$  &  $\chi^2_{r,n,D} = 2.6970$   &  79.5 \% & \\  
$\chi^2_{r,n,T} R^2 = 9.3464 $  &  $\chi^2_{r,n,D} R^2 = 2.7150 $   &  79.5 \% & \\  
$rms_T = 78.2 \mu$K  &  $rms_D = 23.0 \mu$K   &  70.6 \% & (a) \\  
$rms_T = 258 \mu$K  &  $rms_D = 68.3 \mu$K   &  73.5 \% & (b) \\  
$rms_T = 233 \mu$K  &  $rms_D = 46.3 \mu$K   &  80.1 \% & (c) \\  
    & & & \\
    \hline
    \hline
  \end{tabular}
\end{center}


%
%
\bigskip
\bigskip
\begin{center}
  \begin{tabular}{cccc}
    \multicolumn{4}{c}{{\bf Table $9'$}: destriping results; $\alpha=90^{\circ}$; $f_k=10$Hz.} \\
    \hline
    \hline
    & & &\\
    Some global parameters\\
    \hline
    & & & \\
$\Delta T_W = 1.322 $mK  &  $rms_W = 27.18 \mu$K & $R^2 =  1.00666 $ & \% sky = 99.98 \\
Map Res. = 9 & Res. common pix. = 11 & $n_l = 1 $ & pix. in common = $1.62 \times 10^7$ \\
    & & & \\
    \hline
    & & & \\
Before destriping & After destriping & \% improvement & Method \\
    \hline
    & & & \\
$\chi^2_{r,n,T} = 9.2846$  &  $\chi^2_{r,n,D} = 2.7752$   &  78.6 \% & \\  
$\chi^2_{r,n,T} R^2 = 9.3464 $  &  $\chi^2_{r,n,D} R^2 = 2.7937 $   & 78.5 \% & \\  
$rms_T = 78.2 \mu$K  &  $rms_D = 24.2 \mu$K  &  69.0 \% & (a) \\  
$rms_T = 258 \mu$K  &  $rms_D = 70.5 \mu$K   &  72.7 \% & (b) \\  
$rms_T = 233 \mu$K  &  $rms_D = 48.4 \mu$K   &  79.2 \% & (c) \\  
    & & & \\
    \hline
    \hline
  \end{tabular}
\end{center}

\bigskip
\bigskip

\section{Discussion and conclusions} \label{sec_90}

An analytical estimate of the maximum excess noise factor, $F$,
due to the stripes related to $1/f$ effect has been given by
Janssen et al. (1996); they found 
$F \simeq [1+\tau f_k (2 {\rm ln} n_p +0.743)]^{1/2}$.
By thinking $F^2$ as equal to $1 + (\Delta rms / rms_W)^2$, we have 
a fractional additional $rms$ noise 
respect to the $rms$ noise, $rms_W$,  
obtained in the case of pure white noise 
given by
$(\Delta rms / rms_W)^2 = \tau f_k (2 {\rm ln} n_p +0.743)$.
For example for our simulations 
we have $n_p \simeq 2100$, a sampling time $\tau \simeq 28$msec,
corresponding to an angle of $\simeq 10'$ in the sky,
and $f_k=0.05$Hz (or 10 Hz) . With these number we get
$(\Delta rms / rms_W)^2 \simeq 0.022$ (or 4.5) for a pixel of $10' \times 10'$.
Our map pixel is $19.4' \times 19.4'$ (COBE-cube resolution 9); 
therefore  $rms_W^2$ 
reduces by a factor $\simeq 4$ and we expect to have 
an additional $(\Delta rms / rms_W)^2$ per pixel, i.e.
an additional reduced $\chi^2$, of about 0.1 (or 18).
The results shown in our tables are in quite good agreement
(always within a factor 2) with 
these analytical estimates. 
Somewhat larger values may be expected
from the larger observational time toward high ecliptic latitudes,
the consequent reduction of the white noise and the increasing of the relative
weight of the $1/f$ noise.
We stress here that, contrary to the case of pure white noise,  
even for the same scanning strategy 
the final effect of $1/f$ noise may be quite different for different 
simulations: larger or smaller effects can be obtained according
to the (simulated) behaviour of $1/f$ gain fluctuations. Only 
from a very large set of simulations for each considered scanning strategy
we can derive a robust evaluation of the ``averaged'' final effect
of $1/f$ noise (see below). For this reason,
the results shown in our tables have to be considered
as first order estimates of the final $1/f$ noise effect expected
in a given scanning strategy rather than 
detailed predictions. 

Without applying the destriping procedure,
the typical amount of stripes temperatures (i.e. the $rms$ values)
per resolution element (of 
$\simeq 19.4' \times 19.4'$) ranges from few $\mu$K to one or two tens
of $\mu$K 
(according to the method adopted for estimating them) 
and is always significantly less than the
corresponding single beam sensitivity of about $27 \div 30 \mu$K. 

The residual stripes that remain after the reduction through
our destriping code show typical $rms$ temperatures of few 
$\mu$K, roughly independently of the 
adopted scanning strategy. 
The efficiency of the destriping algorithm is quite good
($\sim 20 \% \div 90 \%$, according
to the adopted estimator and depending in part on the scanning
strategy). We stress that in any case, for our kind of radiometers,
the reduced additional noise is of few percent in terms of increased
reduced $\chi^2$ and of few $\mu$K in terms of stripes temperature.
This is not particularly critical. 
Nevertheless it must be compared with
the final Planck sensitivity as it results by combining the data 
from all the receivers at the same frequency (see below). 
 
In the case of much larger $1/f$ contaminations like 
those expected for higher values of the knee frequency
in the case of total power radiometers, 
the final impact on observed sky maps is much higher.
The destriping code allows to reduce a large fraction
of the added noise (see Table 5), nevertheless a significant
increase of the final noise still remains. 
Then, the importance of the reduction via ``hardware'' of the
$1/f$ noise is strongly recommended.

About the dependence of the striping effect 
and of the destriping efficiency on the 
scanning strategy, our
preliminary results suggest the following conclusions.

\begin{itemize}
\item The magnitude of the sensitivity degradation 
due to the stripes
is only weakly dependent on $\alpha$ for undestripped maps as well
as for destripped maps (see Tables 1 $\div$ 3). 

\item The final residual added noise is roughly independent 
of the scanning strategy and on the unreduced added noise,
almost not relevant unreduced contaminations (see Tables 1 $\div$ 4). 

\item Oscillations of the spin axis do not improve
significantly the destriping efficiency (see Table 4). On the contrary
it is well known that oscillations of the spin axis may introduce
further systematic effects related to variations of 
the illumination by the sun and of shielding performances.

\item Our results are of course related to the beam position in the
sky field of view. 
For on-axis beams and for beams 
located at $\phi _B \sim 0^{\circ}$ the case at $\alpha = 90^{\circ}$
is expected to give the worst results, whereas 
even in this case the destriping results are expected to improve
for off-axis beams located at $\phi _B \sim 90^{\circ}$. 

\item 
We find that, in spite of the larger computing time, the
efficiency of the destriping technique does not improve
by using two level constants per scan circle 
(see Tables 6 and 7). Indeed the number of ``physical'' conditions
(the number of pixels in common) do not depend on the 
choosen number of level constants adopted in the destriping
procedure. Then, by searching (in the solution of 
the linear system, see section 5.1, obtained by the 
condition of minimization of $S$)
for a number of unknown larger by a factor two, 
by using the same number of informations, we expect that 
the uncertainty of each unknown will be larger.
We infer that the advantage of using 
more constant per scan circle found by 
Delabrouille (1997)
for a similar knee frequency 
but for the case of thermal drifts (noise spectrum proportional to $1/f^2$)
has to be related to the different kind of noise spectrum.
The noise fluctuations on long timescales will be higher
in the case of $1/f^2$ noise spectrum than in the case  
of $1/f$ noise spectrum; in the former case
the use of more constants per circle may allow a more appropriate
subtraction of the gain fluctuations in each circle, 
whereas in our case this advantage is balanced by the 
increased uncertainty in the determination of the levels
and the global effect results in to a small decreasing
of destriping efficiency.

\item 
The use of more stringent conditions for identifying the pixels
in common does not improve (see Tables 8 and 9) the destriping 
results. This option allows to reach a more accurate superposition
of the pixels used in the destriping; nevertheless the number of couples
decreases about by a factor 4 (or 16) by using pixels 4 (or 16) times
smaller, and so a number of conditions significantly smaller
than that found before enters
in the minimization of $S$ (see section 5.1).

\end{itemize}

Given all these results we believe that a simple scanning strategy
with constant $\alpha$ in the range $85^{\circ} \div 90^{\circ}$  
can offer the advantages of a large (practically full) sky coverage, 
quite good destriping performances together with the minimization
of all the systematic effects related to the variations
of thermal conditions.

\bigskip

We draw here below a brief guide-line for the future simulation work on
this topic.

\begin{itemize}
\item We intend to test different sampling strategies:
for example with a smaller shift (2.5') of the spin axis direction
and a corresponding smaller observation time for spin axis direction (1 hour).
Such a kind of sampling, possibly with
the same ($\sim 8700$, corresponding to 4 samplings
per beam at 100 GHz -- FWHM$\simeq 10'$) number of samplings per scan circle
independently of the frequency, may be the final LFI/Planck sampling.
A strategy of this kind allows to have a larger number of pixels in common 
in the destriping procedure; on the other hand the
sensitivity of the temperature measurements for the different
scan circles degrades.
It may be interesting to investigate pro and contro
of a variance of this kind from the point of view of the $1/f$ noise 
reduction.

\item We may be also interested in an accurate verification of
the validity of the choosen telescope rotation velocity around the spin axis.

\item It may be interesting to apply our codes to higher ratios between
the beam FWHM and the map pixel size, for example by working, at 30 GHz,
with maps at COBE-cube resolution 10. 
Indeed, considering input maps at a higher resolution
roughly tests the importance of assuming
a better efficiency in the data streams deconvolution,
for example by fully exploiting the beam oversampling. 

\item As better approximation, it is interesting
to implement the convolution with the beam for a moving telescope
in the observation simulation code and
to search for robust and fast criteria for establishing in this situation
when it is possible to consider that integrations in different
scan circles can be really referred to the same direction in the sky.
Of course this problem is correlated to the technique adopted for deconvolving
the data streams to obtain observed maps; on the other hand,
the stripes magnitude must be small enough to not significantly
alter the deconvolution procedure.

\item We have found that the noise added by the $1/f$ effect 
is not particularly critical, compared to the single beam 
(white noise) sensitivity. 
Nevertheless it must be compared with
the final Planck sensitivity as it results by combining the data 
from all the receivers at the same frequency. Indeed, we do not expect
that the $1/f$ noise magnitude decreases as the square root
of the number of receivers, as white noise does: by carrying out 
several simulation with the same set of physical parameters 
for the same scanning strategy 
and averaging the corresponding maps, we can address this topic.

\item We intend to apply the methods of inversion of CMB maps
(Muciaccia et al. 1997)
for deriving the angular power spectrum of the observed maps.
By comparing the CMB angular power spectrum obtained in presence
of $1/f$ noise contamination with that derived in the case of pure
white noise (and of course with that of the input map)
it is possible to estimate the $1/f$ noise impact on the extraction
on the key cosmological informations, almost in the case of 
gaussian fluctuations like those expected in inflationary scenarios.
Particular attention has to be attempted for evaluating
the impact on our science in the context of topological defects,
like cosmic strings for example, which introduce non gaussian
features in the power spectrum.
The characteristic geometrical pattern of $1/f$ noise stripes  
(related to the scanning strategy) have to be used for disantangle
between instrumental and cosmological deviations from the gaussianity.

\item Of course, we intend to extend in the next future our analysis to
Planck measurements at higher frequencies.

\end{itemize}

The full success of missions
like Planck and MAP require a good control of all the relevant sources
of systematic effects. Discrete sources above the detection limit
must be carefully removed and accurate models for foregrounds
radiation and anisotropies
(Brandt et al. 1994, Danese et al. 1996, Bouchet et al. 1997, 
Toffolatti et al. 1995, 1997)
are required to keep the sensitivity degradation in the
knowledge of CMB anisotropies below few tens percent (Dodelson 1997).
Optical distortions, which produce a non-symmetric beam response for feed-horns
located away from the centre of the focal plane, introduce other systematic
effects; they must be minimized by optimizing the telescope
and the focal plane assembly design (Mandolesi et al. 1997b).
Thermal drifts (Bersanelli et al. 1996),
which couple to the $1/f$-type noise here discussed, can also
generate stripes in the observed maps; efficient shield
is required together with accurate reduction of sidelobe effects
and optimization of the thermal conditions during the mission.

All in all maximum efforts should be addressed to
optimize the cooling efficiency and
the observational strategy and to improve the
methods for the data analysis in order to reduce the magnitude
of the striping effect and of the other instrumental systematic 
effects.

\bigskip

{\it Acknowledgements} --  We warmly thank M. Seiffert
for useful discussions on Planck LFI receivers and for having
provided us its original IDL code for the generation on $1/f$-type noise,
J. Delabrouille and K. Gorski for useful discussions on simulations 
and destriping techniques during their visits 
in Bologna and P. Natoli and N. Vittorio for having
provided us their code for the generation of CMB anisotropy maps.

\bigskip
\bigskip
\bigskip

\section*{Appendix A: Geometrical transformations between coordinate systems}

Let $\vec i, \vec j, \vec k$ the standard unit vectors in ecliptic 
coordinates and $\vec s$ a unit vector along the satellite
spin axis outward the Sun direction.
Let $i$ the angle 
between $\vec s$ the and $\vec k$ (i.e. the ecliptic colatitude of $\vec s$) 
and $\phi$ the angle between $\vec i$ and $\vec s$ (i.e. the ecliptic 
longitude of $\vec s$) . 
For general scanning strategies $i$ will
be described by a, possibly not constant, function $i=i(\phi)$.
Then 
$\vec s = {\rm sin} \, i {\rm cos} \, \phi \, \, \vec i 
+ {\rm sin} \, i {\rm sin} \, \phi \, \, \vec j 
+ {\rm cos} \, i \, \, \vec k$.
Let $\vec i' = \vec s$ and $\vec k'$ a unit vector hortogonal to 
$\vec s$ on the plane identified by the vectors $\vec k$ and $\vec s$, 
namely 
$\vec k' = -{\rm cos} \, i {\rm cos} \, \phi \, \, \vec i 
- {\rm cos} \, i {\rm sin} \, \phi \, \, \vec j + {\rm sin} \, i \, \, \vec k$.
Let $\vec j' = \vec k' \wedge \vec i'$ (here $\wedge$ indicates the vector product).
Let $\vec p$ the unit vector that identifies the pointing direction of the 
telescope optical axis. In the reference 
$\vec i', \vec j', \vec k'$ the vector $\vec p$ can be defined
by two angles: the angle $\alpha$ from $\vec s$ ($\alpha=70\deg$ for
the Phase A study, Bersanelli et al. 1996) and the angle, $\psi$, 
between its projection on the plane identified by  $\vec j', \vec k'$ 
and $\vec k'$, with the convention
$\vec p = {\rm cos} \, \alpha \, \, \vec i' 
+ {\rm sin} \, \alpha {\rm sin} \, \psi \, \, \vec j' 
+ {\rm sin} \, \alpha {\rm cos} \, \psi \, \, \vec k'$.
Given $\vec i', \vec j', \vec k'$ in terms of $\vec i, \vec j, \vec k$ 
it is easily to derive $\vec p$ in the same basis.
We choose two coordinates $x$ and $y$ on the plane tangent to the celestial
sphere in the telescope optical axis direction, $\vec p$, 
with unit vector $\vec u$ and $\vec v$ respectively;
we choose the $x$ axis according
to the condition that the unit vector $\vec u$ points always toward the
satellite spin axis; indeed, for standard Planck observational strategy,
this condition is preserved as the telescope scans different sky regions.
With this choice of reference frame, we have that
$\vec v = \vec p \wedge \vec s / |\vec p \wedge \vec s|$ and
$\vec u = \vec v \wedge \vec p / |\vec v \wedge \vec p|$.
In general, the coordinates $(x_0,y_0)$ of the beam centre
in a (``satellite'') reference $x_T, y_T, z_T$, corresponding to the unit vectors
$\vec u, \vec v, \vec p$, 
can be identified by two angles; we use here the colatitude $\theta_B$ and
the longitude $\phi_B$ in this reference.
Finally, the pointing direction of this generic (on-axis or off-axis) beam 
is given by the unit vector 
$\vec B = {\rm cos} \, \theta_B \, \, \vec p 
+ {\rm cos} \, \phi_B {\rm sin} \, \theta_B \, \, \vec u 
+ {\rm sin} \, \phi_B {\rm sin} \, \theta_B \, \, \vec v$.

For sake of illustration, we consider here briefly 
three different kinds of scanning strategies
(the first two options have been considered
in the present simulations).

\begin{itemize}

\item {\it Spin axis always on the ecliptic plane.}

\noindent
In this simple case we have $i=90^{\circ}$ and 
$\vec s = {\rm cos} \, \phi \, \, \vec i + {\rm sin} \, \phi \, \, \vec j$.

Given the angle $\alpha$, 
we need to give the time dependences, 
$\phi = \phi (t)$ and $\psi = \psi(t)$, 
of the spin axis longitude and of the telescope projection
to fully determine the  scanning strategy.
The reference case is to change $\phi$ of a certain 
angle $\Delta \phi$ ($5'$, for example) after a given
time interval (2 hours, for example) and to choose
a given spin frequency (1 r.p.m., for example) 
of the continuous rotation of $\psi$.

\item {\it Sinusoidal oscillations of the spin axis.}

\noindent
In this case we need also to define the amplitude of the
oscillations, $\delta$ ($10^{\circ}$, for example),
and the number of complete oscillations, $n_{osc}$
(10 oscillations for 360 days, for example), per a complete
rotation of the spin axis over the ecliptic
(in 360 days, for example).  
Then, by choosing $i=0$ when $\phi=0$, we have 
simply $i=\delta {\rm sin} (n_{osc} \phi)$. 

\item {\it Precession of the spin axis.}

This case is just a little more complicate.
We can consider a further unit vector $\vec f$ 
which moves always on the ecliptic plane;
its ecliptic longitude, $\eta = \eta (t)$
defines it: $\vec f = {\rm cos} \, \eta \, \, \vec i + {\rm sin} \, \eta \, \, \vec j$.
Let $\vec j'' = -{\rm sin} \, \eta \, \, \vec i + {\rm cos} \, \eta \, \, \vec j$.
The satellite spin axis $\vec s$ precedes around $\vec f$
(we can consider for example again 10 precessions for 360 days, 
per a complete rotation of the axis $\vec f$ over 
the ecliptic plane in 360 days).  
Let $\xi$ the angle between its projection on 
the plane identified by $\vec j''$ and $\vec k$ and
the vector $\vec k$ and $\delta$ the angle between
$\vec f$ and $\vec s$ ($10^{\circ}$, for example). 
Then, the relation
$\vec s = {\rm cos} \, \delta \, \, \vec f 
+ {\rm sin} \, \delta {\rm cos} \, \xi \, \, \vec k 
+ {\rm sin} \, \delta {\rm sin} \, \xi \, \, \vec j''$ easily gives
the colatitude $i$ and the longitude $\phi$ of spin axis.

\end{itemize}

\section*{Appendix B: System creation and solution with low memory usage}

As discussed in section~\ref{sec_50_20}, the creation of the linear system
requires a large amount of memory, usually much more than the available RAM.
This problem can be avoided by taking advantage of disk buffers, essentially
by splitting a large matrix into smaller blocks and creating it a block at a
time. Our strategy to do this is very simple:

\begin{enumerate}
\item a memory buffer is created, large enough to keep $L$ lines;

\item the algorithm described in section~\ref{sec_50_10} is performed: for
each couple $\pi$ of pixel in common between two scan circles $i$ and $j$,
the quantities $\chi_\pi$ and $\tau_\pi$ are evaluated;

\item if $i$ is in the range $[0, \ldots, L-1]$, then equations $(17)$, 
$(18)$ and $(21)$ are applied; if $j$ is in the range $[0, \ldots, L-1]$, 
then equations $(19)$, $(20)$ and $(22)$ are applied;

\item after all the couples have been evaluated, the memory buffer is
saved in a file; steps $2-4$ are then repeated for $i$ and $j$ in the
range $[L, \ldots, 2L-1]$, and then for $i$ and $j$ in $[2L, \ldots, 3L-1]$,
and so on until all the $N$ matrix lines have been created.
\end{enumerate}

At the end of this loop, the linear system has been created and stored in 
a file in binary format; the coefficients are organized in a matrix of $N$
rows and $N+1$ columns, where the last column contains the known terms. 

It is easy to see that this strategy saves memory space but increases
considerably CPU time, because time required to create the whole matrix
is directly proportional to the number of pieces $N/L$ into which the
linear system is divided. Due to the fact that program execution time
is essentially dominated by the routines which solve the system, while creation 
time is negligible for the present practical purpose, we didn't bother to
improve our code.

A very different situation occurs for system solution, because both
a large amount of memory and a lot of computation time are usually
required. While time efficiency can't be usually improved (except in 
a few particular cases, when the matrix which defines the system 
has special symmetry properties and a lot of zero coefficients), 
memory requirements can be considerably reduced by taking advantages of
disk swapping. However, a special care is required when choosing the
strategy for disk operations, because disk time can easily blow up and
overtake CPU time. Our algorithm behaves as follows:

\begin{enumerate}
\item as for system creation, a memory buffer is allocated, large enough
  to keep $L$ lines;

\item then the first $L$ lines ($0, \dots, L-1$) are loaded in memory, and
  complete Gauss elimination is performed on them: each line is reduced by
  the preceding lines and is used to reduce the following lines;

\item each of the remaining $N-L$ lines is sequentially load into memory, and
  reduced by \emph{each of} the $L$ lines stored in the memory buffer. In
  this way, we perform $L$ steps of the Gauss elimination algorithm with a
  \emph{single} disk operation;

\item the memory buffer is flushed, and the next $L$ lines ($L, \dots, 2L-1$) 
  are loaded into memory, reduced by one another and used to sequentially 
  reduce all the remaining lines of the system. Steps 3-4 are repeated until
  the original matrix has been completely reduced.

\end{enumerate}

As can be seen, our algorithm differs from the ``standard'' version only in
the \emph{order} in which the elimination is performed; in this way, we can
limit memory requirements without increasing total solution time. It is
easy to see that the total time spent for disk operation is:
\begin{equation}
  t_\mathrm{disk} = \beta \frac N2 \left( \frac NL + 1 \right) (N+1) \approx 
  \beta_\mathrm{disk} \frac {N^3}L
\end{equation}

CPU time scales as the cube of the linear size of the system:
\begin{equation}
  t_\mathrm{cpu} = \beta_\mathrm{cpu} N^3
\end{equation}

So we conclude that the ratio $\eta$ between disk time and CPU time is
independent from the size $N$ of the linear system, and only depend on the
total number of lines $L$ then we can hold simultaneously in memory:
\begin{equation}
  \eta = \frac{t_{disk}}{t_{cpu}} = \frac{\beta_{disk}}{\beta_{cpu} \cdot L}
\end{equation}

Of course, the memory necessary to hold a line is proportional to the line
length, so we need a larger amount of memory to solve a larger system;
however, the size of the buffer is \emph{linear} in system size $N$, while the
memory required to hold all the coefficients simultaneously is
\emph{quadratic} in $N$.



\bigskip
\bigskip

\centerline{\bf REFERENCES}

\bigskip

\noindent
Bersanelli M., Mandolesi N., Weinreb S., Ambrosini R. \& Smoot G.F., 1995,
Int. Rep. ITESRE 177/1995 -- COBRAS memo n.5

\noindent
Bersanelli M. et al., 1996. ESA, COBRAS/SAMBA Report on the Phase A Study,
D/SCI(96)3

\noindent
Blum E.J., 1959, Annales d'Astrophysique, 22-2, 140

\noindent
Bouchet F. et al., 1997, in proceedings of {\it The XVIth Moriond 
Astrophysics Meeting}, Les Arcs, Savoie, France, 16-23 March 1996.
 
\noindent
Brandt W.N. et al., 1994, ApJ 424, 1

\noindent
Burigana C. et al., 1997a, Int. Rep. ITeSRE/CNR 186/1997

\noindent
Burigana C. et al., 1997b, in proceedings of {\it Particle Physics
and Early Universe Conference}, Cambridge 7-11 April 1997,
http://www.mrao.cam.ac.uk/ppeuc/proceedings/

\noindent
Burigana C. et al., 1997c, A\&A, submitted

\noindent
Colvin R.S., 1961, Ph.D. thesis, Stanford University

\noindent
Danese L., Toffolatti L., Franceschini A., Bersanelli M.
\& Mandolesi N. 1996, Astroph. Lett \& comm, 33, 257.

\noindent
Delabrouille J., 1997, A\&A, submitted

\noindent
Dodelson S., 1997, ApJ 482, 577



\noindent
Gaier T., 1997, private communication


\noindent
Janssen M.A. et al., 1996, Astrophys. J. Lett., submitted

\noindent
Jarosik N.C., 1996, IEEE Trans. Microwave Theory Tech., 44, 193


\noindent
Mandolesi N. et al., 1997a, in proceedings of {\it Particle Physics
    and Early Universe Conference}, Cambridge 7-11 April 1997,
  http://www.mrao.cam.ac.uk/ppeuc/proceedings/

\noindent
Mandolesi N. et al., 1997b, Int. Rep. ITeSRE/CNR 198/1997

\noindent
Muciaccia P.F. et al., 1997, preprint astro-ph/9703084




\noindent
Pospieszalsky, 1989, MTT Sep, p. 1340

\noindent
Press W.H. et al., 1992, {\rm ``Numerical Recipes in Fortran''},
Cambridge University Press

\noindent
Seiffert M. et al., 1996, Rev. Sci. Instrum., submitted

\noindent
Strang G., 1976, {\rm ``Linear Algebra and Its Applications''},
Academic Press, Inc.

\noindent
Toffolatti L. et al., 1995, Astro. Lett. \& Comm. 32, 125

\noindent
Toffolatti L. et al., 1997, MNRAS, submitted


\noindent
Weinreb S., 1997, private communication


\noindent
Wollack E.J., 1995, Rev. Sci. Instrum., 66, 4305



\newpage
\bigskip
\bigskip
\bigskip

\centerline{\bf FIGURE CAPTIONS}

\bigskip
\bigskip

\noindent
{\bf Figure 1:} 
The simulated (input) map of CMB anisotropies 
(CDM model, the dipole term is neglected).
Galactic coordinates have been used for the plot.

\bigskip

\noindent
{\bf Figure 2:} 
Schematic representation of the observational geometry.

\bigskip

\noindent
{\bf Figure 3:} 
The unreduced simulated noise map 
(white plus $1/f$ noise) for the simulation
with $\alpha = 85^{\circ}$.
Note the two small circular regions close to the ecliptic poles 
that are not observed by the considered off-axis beam; 
for graphic purposes we have filled them
with a random noise distribution with variance given by the noise variance of the 
observed pixels. Note also the elongated sky region with noise significantly
larger than the average, which corresponds to the sky regions that are 
observed a single time only, due to the choosen value of $\alpha$.
(Galactic coordinates have been used for the plot).

\bigskip

\noindent
{\bf Figure 4:} 
The reduced simulated noise map 
(white plus ``reduced'' $1/f$ noise) for the simulation
with $\alpha = 85^{\circ}$.
Note again the two small circular regions close to the ecliptic poles 
filled with a random noise distribution.
We note that the 
the elongated sky region with noise significantly
larger than the average disappears as result of the destriping procedure.
Also the stripes in the sky became much less evident.
(Galactic coordinates have been used for the plot).

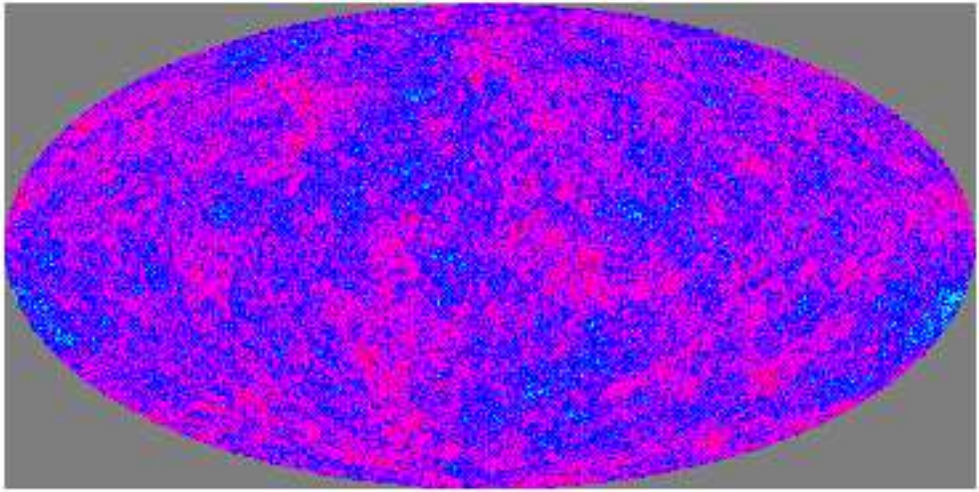
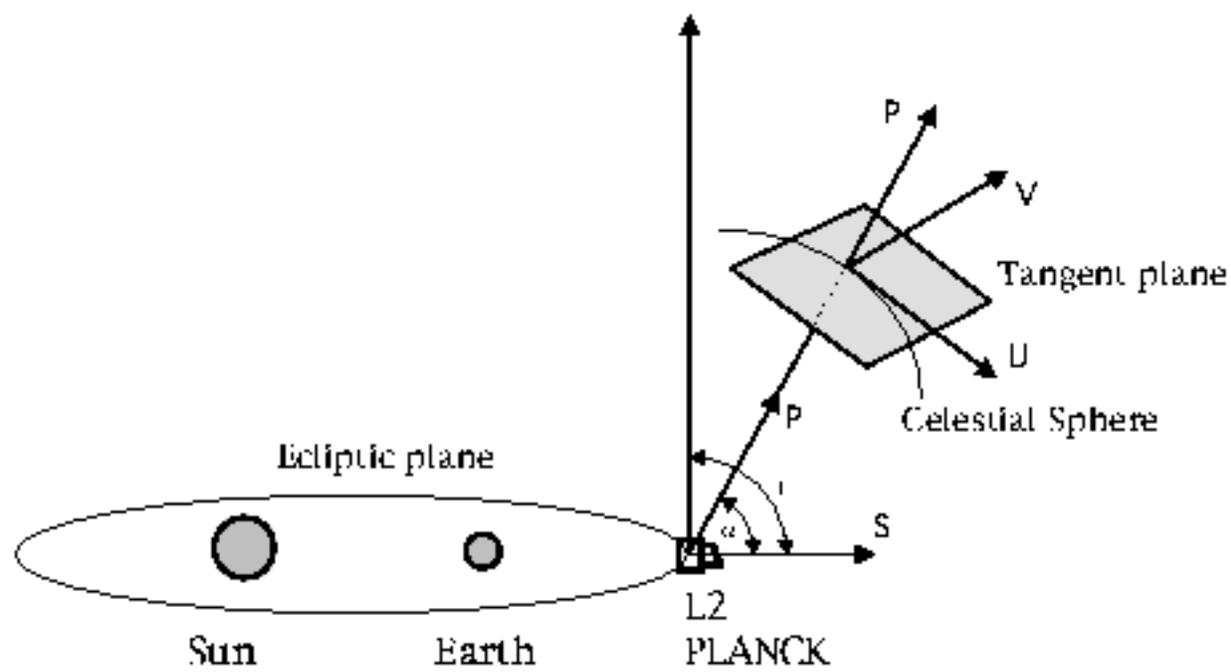
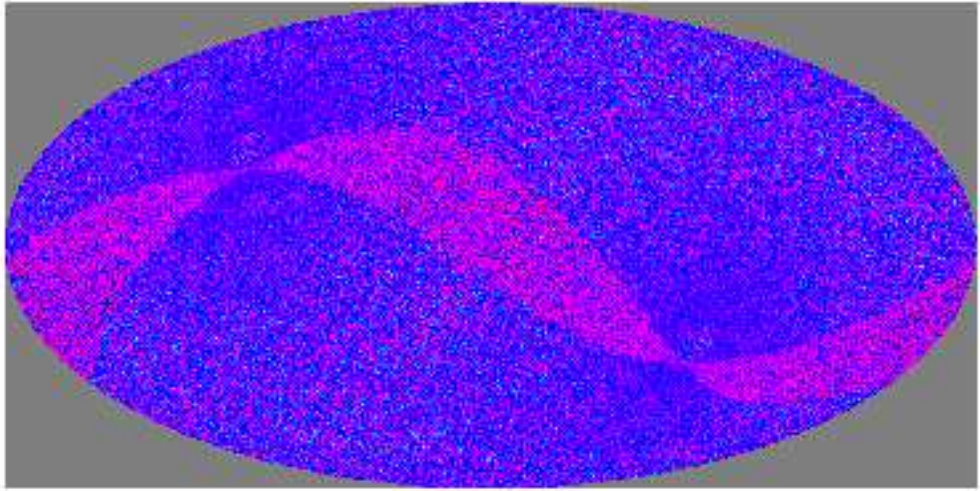
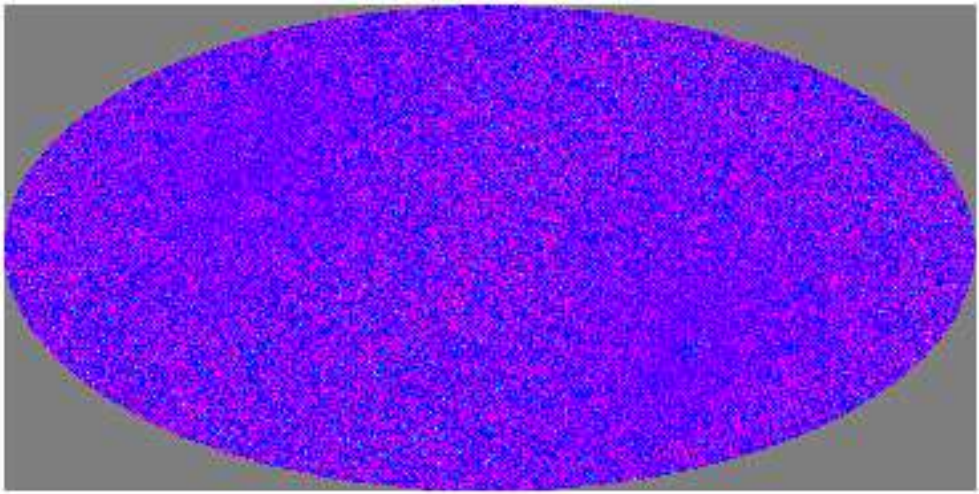
\end{document}